\documentclass{article}
\usepackage{arxiv}

\usepackage[utf8]{inputenc} 
\usepackage[T1]{fontenc}    
\usepackage{hyperref}       
\usepackage{url}            
\usepackage{booktabs}       
\usepackage{amsfonts}       
\usepackage{nicefrac}       
\usepackage{microtype}      
\usepackage{lipsum}
\usepackage{float}
\usepackage{graphicx}
\usepackage{multirow}
\usepackage{array}
\usepackage{comment}
\usepackage{longtable}
\usepackage{ragged2e}
\usepackage{subfig}
\usepackage[table]{xcolor}  
\usepackage{geometry}
\usepackage{multicol}
\usepackage{amsmath}
\usepackage{pifont}
\title{Diffusion-Based Approaches in Medical Image Generation and Analysis}

\author{
  Abdullah al Nomaan Nafi
 \\
  Department of Biomedical Engineering \\
  Islamic University \\ Kushtia 7003, Bangladesh\\
  \texttt{nomaan.ict.iu@gmail.com} \\
  \And
    Md. Alamgir Hossain
 \\
  Department of Biomedical Engineering \\
  Islamic University \\ Kushtia 7003, Bangladesh\\
  \texttt{hossain@iu.ac.bd } \\
  \And
  Rakib Hossain Rifat, Md Mahabub Uz Zaman \\
  Dept. of Computer Science \\
  Texas Tech University \\
   Lubbock, TX 79409\\
  \texttt{\{rrifat, m.zaman\}@ttu.edu} \\
  \And
  Md Manjurul Ahsan \\
  Department of Industrial and Systems Engineering\\
  University of Oklahoma\\
  Norman, Oklahoma-73071 \\
  \texttt{ahsan@ou.edu} \\
   \And
  Shivakumar Raman \\
  Department of Industrial and Systems Engineering\\
  University of Oklahoma\\
  Norman, Oklahoma-73071\\
  \texttt{raman@ou.edu}  
}

\begin{document}
\maketitle

\begin{abstract}

Data scarcity in medical imaging poses significant challenges due to privacy concerns. Diffusion models, a recent generative modeling technique, offer a potential solution by generating synthetic and realistic data. However, questions remain about the performance of convolutional neural network (CNN) models on original and synthetic datasets. If diffusion-generated samples can help CNN models perform comparably to those trained on original datasets, reliance on patient-specific data for training CNNs might be reduced. In this study, we investigated the effectiveness of diffusion models for generating synthetic medical images to train CNNs in three domains: Brain Tumor MRI, Acute Lymphoblastic Leukemia (ALL), and SARS-CoV-2 CT scans. A diffusion model was trained to generate synthetic datasets for each domain. Pre-trained CNN architectures were then trained on these synthetic datasets and evaluated on unseen real data. All three datasets achieved promising classification performance using CNNs trained on synthetic data. Local Interpretable Model-Agnostic Explanations (LIME) analysis revealed that the models focused on relevant image features for classification. This study demonstrates the potential of diffusion models to generate synthetic medical images for training CNNs in medical image analysis.

\end{abstract}



\section{Introduction}
Medical image classification holds immense promise for revolutionizing healthcare  \cite{litjens2017survey, yu2016automated}. Its potential applications span disease diagnosis, treatment planning, and patient prognosis, offering tools to improve patient outcomes and advance medical understanding. However, this potential is hindered by a significant challenge: data scarcity.

Limited availability of labeled data plagues medical image classification due to several factors. Ethical considerations around patient privacy and data security often restrict data collection and sharing. The cost of acquiring medical imaging equipment and the expertise needed for data interpretation further limit data generation. Additionally, the rarity of certain diseases creates an imbalance in class distribution, where some classes have significantly fewer data points compared to others.

These challenges are further amplified by the complexity and cost of labeling medical images. Annotations often require specialized knowledge from physicians, making the process time-consuming and expensive. Furthermore, data variability adds another layer of complexity. Diverse acquisition factors, such as different imaging equipment and protocols, can lead to significant variations in image quality and appearance. Similarly, patient factors like age, ethnicity, and individual anatomy can cause the same disease to manifest differently in different individuals, creating inconsistencies within the data. Finally, privacy and security concerns surrounding patient data restrict data sharing and collaboration, hindering the creation of larger datasets crucial for training robust models.

In medical image analysis, deep learning models, particularly Convolutional Neural Networks (CNNs), have achieved remarkable success in various tasks like tumor segmentation, disease classification, and anomaly detection \cite{litjens2017survey, yu2016automated}. However, a significant challenge hindering the development of robust CNN models lies in the limited availability of high-quality, labeled medical image datasets. This scarcity of data can lead to model overfitting and hinder generalizability to unseen data \cite{avvakumov2021envy}.

Synthetic data generation using diffusion models (DMs) offers a promising solution to address data scarcity in medical image analysis \cite{pinaya2022brain, ozbey2023unsupervised, chung2022score, wolleb2022diffusion}. DMs are generative models trained to progressively remove noise added to real data points, ultimately enabling them to create realistic and statistically similar synthetic samples \cite{zamzmi2020unified}. This methodology allows for the creation of large-scale synthetic medical image datasets that can be leveraged to train CNNs effectively.

This study explores the potential of using synthetic data generated by a diffusion model for training CNNs in various medical image analysis tasks. We investigate the effectiveness of this approach in overcoming data limitations and achieving robust and generalizable CNN models by experimenting with 3 distinct datasets and 8 distinct model architectures.

\subsection{Motivation}

Medical image analysis plays a critical role in healthcare, enabling applications like disease diagnosis, treatment planning, and surgical guidance \cite{litjens2017survey}. Deep learning models, particularly Convolutional Neural Networks (CNNs), have demonstrated remarkable success in these tasks, achieving high accuracy in tumor segmentation, disease classification, and anomaly detection \cite{litjens2017survey}. However, a significant challenge hindering the development of robust CNN models lies in the limited availability of high-quality, labeled medical image datasets.
\\[1\baselineskip]
This data scarcity arises from several factors. Privacy concerns and patient confidentiality make it difficult to collect large datasets of medical images \cite{yu2016automated, avvakumov2021envy}. Additionally, the annotation process, where experts label specific features within the images, is expensive and time-consuming. Limited data availability can lead to issues like overfitting, where the model performs well on the training data but fails to generalize effectively to unseen data \cite{yu2016automated, avvakumov2021envy}.
\\[1\baselineskip]
Furthermore, existing medical image datasets may exhibit biases due to factors like patient demographics, healthcare access disparities, or limitations in data collection procedures. These biases can negatively impact the performance and generalizability of trained CNN models \cite{zamzmi2020unified}.
\\[1\baselineskip]
Several approaches have been explored to address data scarcity and bias in medical image analysis. Data augmentation techniques, like random flipping, rotations, and scaling, can be applied to existing datasets to artificially increase the data size and improve model generalizability \cite{goodfellow2014generative}. However, these techniques only generate variations of existing data and may not address the underlying issue of limited unique samples. Transfer learning, where pre-trained CNN models are fine-tuned on smaller medical image datasets, can leverage existing knowledge but might not fully capture the specific characteristics of medical images \cite{isola2017image}.
\\[1\baselineskip]
Synthetic data generation using generative models offers a promising solution to address data scarcity and bias in medical image analysis. These models can learn the underlying statistical distribution of real data and produce realistic, synthetic samples that share similar characteristics with the original data \cite{tang2019ct}. This approach can significantly increase the size and diversity of datasets used to train CNN models.
\\[1\baselineskip]
Diffusion models (DMs) are a class of generative models that have recently gained significant attention for their ability to generate high-fidelity synthetic data \cite{popescu2021retinal}. DMs work by progressively removing noise added to real data points. By learning this noise removal process in reverse, DMs can eventually create new, realistic samples that resemble the original data \cite{popescu2021retinal}. This approach has the potential to address data scarcity and bias limitations by generating a larger and more representative dataset for training CNN models in medical image analysis tasks.
\\[1\baselineskip]
Motivated by the limitations of traditional methods and the promising potential of diffusion models, this study aims to investigate the effectiveness of using diffusion models to generate synthetic medical images for training CNN models in medical image analysis tasks. We hypothesize that diffusion models can be used to create high-quality synthetic medical images that can improve the performance and generalizability of CNN models compared to models trained on limited real data or data augmented with traditional techniques.
\subsection{Contribution}
This study investigates the potential of diffusion models (DMs) for generating synthetic medical image data to train Convolutional Neural Networks (CNNs) in various medical image analysis tasks. Our work makes the following key contributions:
\begin{itemize}
\item \textbf{Leveraging Diffusion Models for Synthetic Medical Image Generation:} We demonstrate the effectiveness of using a diffusion model to create large-scale datasets of synthetic medical images. This approach addresses the challenge of data scarcity often encountered in medical image analysis tasks.

\item \textbf{Improved CNN Training with Synthetic Data:} We show that training CNNs on synthetic data generated by the diffusion model leads to robust and generalizable models. This approach has the potential to overcome limitations associated with training on smaller real-world medical image datasets.

\item \textbf{Exploration of Multiple CNN Architectures:} We explore the performance of eight distinct CNN architectures when trained on synthetic data. This investigation helps identify architectures well-suited for specific medical image analysis tasks and synthetic data characteristics.

\item \textbf{Incorporation of Explainable AI (XAI):} We employ LIME (Local Interpretable Model-agnostic Explanations) to gain insights into the decision-making processes of the trained CNN models. This analysis enhances interpretability and trust in the model's predictions for medical image analysis tasks.
\end{itemize}
Overall, this study contributes to the advancement of deep learning in medical image analysis by exploring the use of diffusion models for synthetic data generation and its impact on CNN training. The findings offer valuable insights for researchers and practitioners seeking to leverage deep learning models for effective medical image analysis tasks, even in scenarios with limited real-world data availability.
\section{Literature Review}
Before diffusion models became popular in medical image analysis or in mainstream computer vision, GANs \cite{goodfellow2014generative} were the most popular image generation methods. Developed to perform conditional natural image generation, Pix2PixGAN \cite{isola2017image} was adapted to medical imaging and several researchers have shown its usefulness in such tasks \cite{tang2019ct, popescu2021retinal, aljohani2022generating, sun2022pix2pix}. Zhu et al. \cite{zhu2017unpaired} proposed CycleGAN to perform conditional image-to-image translation between two domains using unpaired images, and the model has also been extensively used in medical imaging. Du et al. \cite{du2018reduction} made use of CycleGAN in CT image artifact reduction. Yang et al. \cite{yang2018unpaired} used a structure-constrained CycleGAN to perform unpaired MRI-to-CT brain image generation. Liu et al. \cite{liu2021ct} utilized multi-cycle GAN to synthesize CT images from MRI for head-neck radiotherapy. Harms et al. \cite{harms2019paired} applied CycleGAN to image correction for cone-beam computed tomography (CBCT). Karras et al. \cite{karras2019style} proposed StyleGAN, which has an automatically learned, unsupervised separation of high-level attributes and stochastic variation in the generated images, enabling easier control of the image synthesis process. 
\\[1\baselineskip]
Fetty et al. \cite{fetty2020latent} manipulated the latent space for high-resolution medical image synthesis via StyleGAN. Su et al. \cite{su2020pre} performed data augmentation for brain CT motion artifact detection using StyleGAN. Hong et al. \cite{hong20213d} introduced 3D StyleGAN for volumetric medical image generation. Other GAN-based methods have also been proposed for medical imaging. Progressive GAN \cite{mahapatra2019image} was used to perform medical image super-resolution. Upadhyay et al. \cite{upadhyay2021uncertainty} extended the model by utilizing uncertainty estimation to focus more on the uncertain regions during image generation. Armanious et al. \cite{armanious2020medgan} proposed MedGAN, specific to medical image domain adaptation, which captured the high and low-frequency components of the desired target modality.
\\[1\baselineskip]
Apart from GANs, other generative models, including VAEs and NFs, are also popular in image generation. The VAE was introduced by Kingma and Welling \cite{kingma2013auto}, and it has been the basis for a variety of methods for image generation. Vahdat and Kautz \cite{vahdat2020nvae} developed Nouveau VAE (NVAE), a hierarchical VAE that is able to generate highly realistic images. Hung et al. \cite{hung2021hierarchical} adapted some of the features from NVAE into their hierarchical conditional VAE for ultrasound image inpainting. Cui et al. \cite{cui2021pet} adopted NVAE in positron emission tomography (PET) scan image denoising and uncertainty estimation. As for the NF models, Grover et al. \cite{grover2020alignflow} proposed AlignFlow based on a similar concept with NF models instead of GANs. Bui et al. \cite{bui2020flow} extended AlignFlow into medical imaging for Unpaired multi-contrast MRI conditional image generation. Wang et al. \cite{wang2021harmonization} and Beizaee et al. \cite{hung2023med} applied NF to medical image harmonization.
\\[1\baselineskip]
In recent years, diffusion models have become the most dominant algorithm in image generation due to their ability to generate realistic images. On natural images, diffusion models have achieved SOTA results in unconditional image generation by outperforming their GAN counterparts \cite{ho2020denoising, dhariwal2021diffusion}. Diffusion models have achieved outstanding performance in tasks such as super-resolution \cite{saharia2022image, kadkhodaie2020solving}, image editing \cite{meng2021sdedit, sinha2021d2c}, and unpaired conditional image generation \cite{sasaki2021unit}, and they have attained SOTA performance in conditional image generation \cite{saharia2022palette}. In medical imaging, unsupervised anomaly detection is an important application of unconditional diffusion models \cite{pinaya2022brain, wolleb2022diffusion, behrendt2024patched, behrendt2024patched}. Image segmentation is a popular application of conditional diffusion models, where the image to be segmented is used as the condition \cite{wolleb2022diffusion, rahman2023ambiguous, wu2024medsegdiff, zbinden2023stochastic, chen2023berdiff}. Diffusion models have also been widely applied to accelerating MRI reconstruction \cite{chung2022score, cao2024high, luo2023bayesian}. Özbey et al. \cite{ozbey2023unsupervised} used GANs to shorten the denoising process in diffusion models for medical imaging.

\section{Basics of Diffusion Model}
Diffusion models are a cutting-edge class of generative models that have been demonstrated to be highly effective in learning complex data distributions. They are a relatively new addition to the generative learning landscape but have shown to be useful in various applications. In this section, we take an in-depth look at the theory of diffusion models. We begin by discussing the position of diffusion models within the broader generative learning landscape and provide a new perspective on how they compare to other generative models. We further classify diffusion models into two main perspectives: the Variational Perspective and the Score Perspective. We delve into their details and highlight the DDPMs. Ultimately, we provide a comprehensive understanding of the underlying theory behind this method.
\label{c3}
\subsection{Where do diffusion models fit the generative learning landscape?
}
\begin{figure}
    \centering
    \includegraphics[width= \linewidth,]{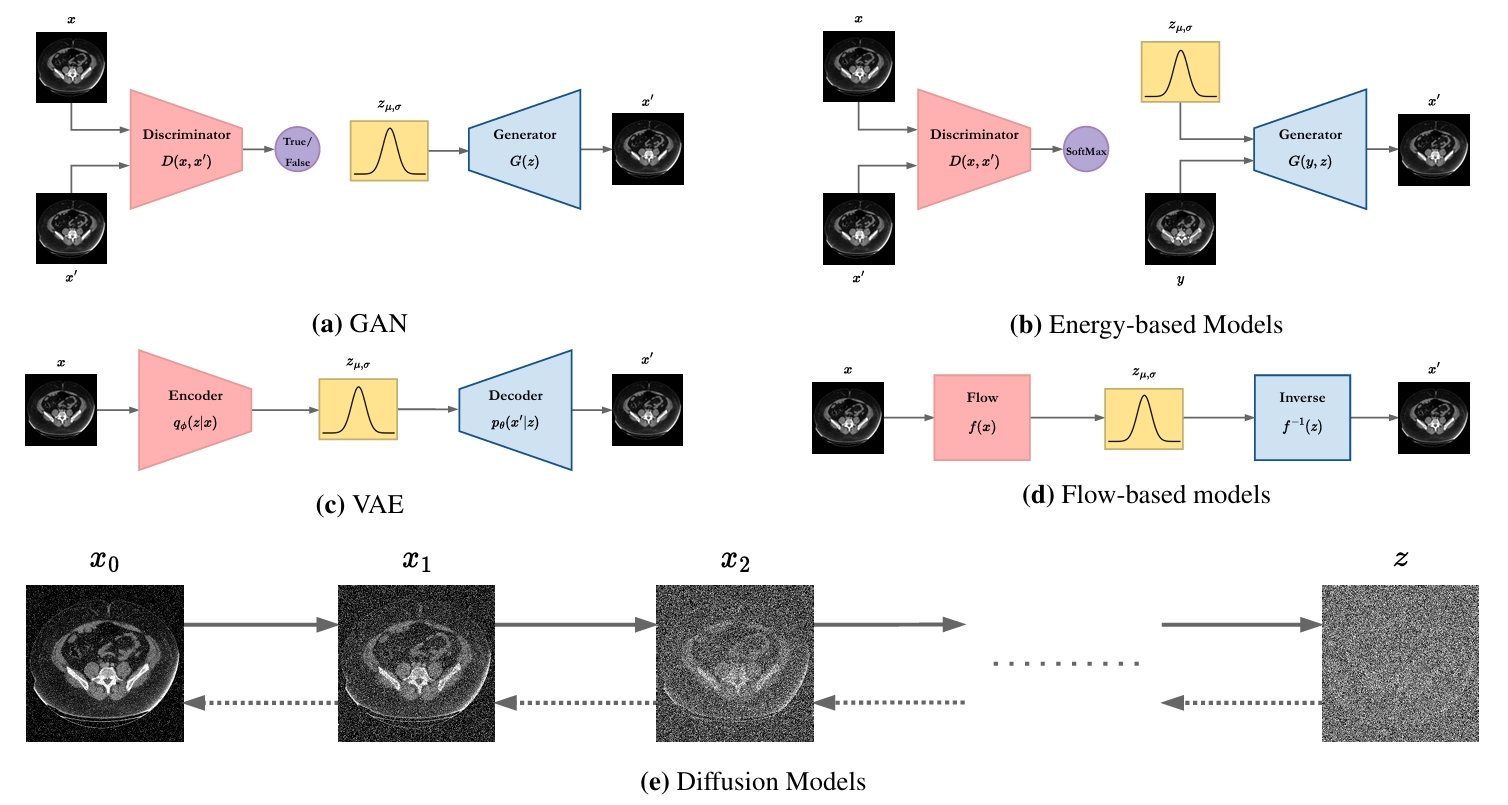}
    \caption{ This figure showcases different generative models and provides an overview of their underlying principles \cite{kazerouni2023diffusion} }
     \label{fig: 3.1}
\end{figure}

\begin{figure}
    \centering
    \includegraphics[width= 100mm, height=65mm]{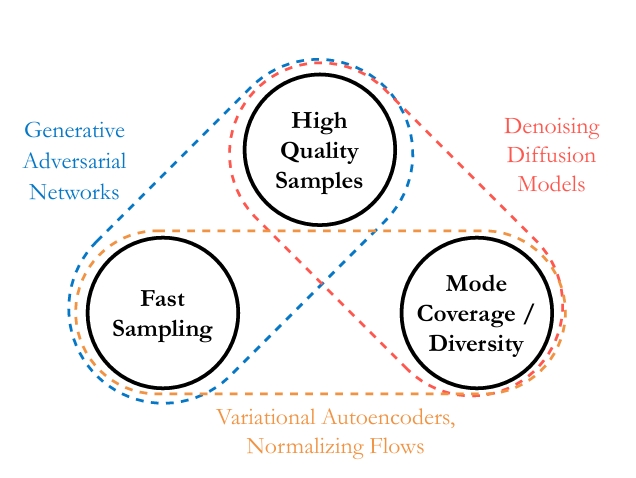}
    \caption{Generative learning trilemma \cite{kazerouni2023diffusion}}
     \label{fig: 3.2}
\end{figure}

Following the remarkable surge of available datasets, as well as advances in general deep learning architectures, there has been a revolutionary paradigm shift in generative modeling. Specifically, the three mainstream generative frameworks include, namely, GANs \cite{popescu2021retinal}, VAEs \cite{xie2022measurement, kingma2013auto}, and normalizing flows \cite{dinh2016density} (see Figure \ref{fig: 3.1}). Generative models typically entail key requirements to be adopted in real-world problems. These requirements include (i) high-quality sampling, (ii) mode coverage and sample diversity, and (iii) fast execution time and computationally inexpensive sampling \cite{xiao2021tackling} (see Figure \ref{fig: 3.2}). Generative models often make accommodations between these criteria. Specifically, GANs are capable of generating high-quality samples rapidly, but they have poor mode coverage and are prone to lack sampling diversity. Conversely, VAEs and normalizing flows suffer from the intrinsic property of low sample quality despite being witnessed in covering data modes. GANs consist of two models: a generator and a critic (discriminator), which compete with each other while the discriminator, which is typically a binary classifier, estimates the probability of a given sample coming from the real dataset. It works as a critic and is optimized to recognize the synthetic samples from the real ones. A common concern with GANs is their training dynamics which have been recognized as being unstable, resulting in deficiencies such as mode collapse, vanishing gradients, and convergence \cite{wiatrak2019stabilizing}. Therefore, an immense interest has also influenced the research direction of GAN to propose more efficient variants \cite{miyato2018spectral, motwani2020novel}. VAEs optimize the log-likelihood of the data by maximizing the evidence lower bound (ELBO). Despite the remarkable achievements, the behavior of VAEs is still far from satisfactory due to some theoretical and practical challenges such as balancing issues \cite{davidson2018hyperspherical} and the variable collapse phenomenon \cite{asperti2019variational}. A flow-based generative model is constructed by a sequence of invertible transformations. Specifically, a normalizing flow transforms a simple distribution into a complex one by applying a sequence of invertible transformation functions where one can obtain the desired probability distribution for the final target variable using a change of variables theorem. Unlike GANs and VAEs, these models explicitly learn the data distribution; therefore, their loss function is simply the negative log-likelihood \cite{montecchiani2022survey}. Despite being feasibly designed, these generative models have their specific drawbacks. Since the Likelihood-based method has to construct a normalized probability model, a specific type of architecture must be used (Autoregressive Model, Flow Model), or in the case of VAE, an alternative Loss such as ELBO is not calculated directly for the generated probability distribution. In contrast, the learning process of GANs is inherently unstable due to the nature of the adversarial loss of the GAN. Recently, diffusion models \cite{saharia2022image, kadkhodaie2020solving} have emerged as powerful generative models, showcasing one of the leading topics in computer vision so that researchers and practitioners alike may find it challenging to keep pace with the rate of innovation. Diffusion models are a powerful class of probabilistic generative models that are used to learn complex data distributions. These models accomplish this by utilizing two key stages: the forward diffusion process and the reverse diffusion process. The forward diffusion process adds noise to the input data, gradually increasing the noise level until the data is transformed into pure Gaussian noise. This process systematically perturbs the structure of the data distribution. The reverse diffusion process, also known as denoising, is then applied to recover the original structure of the data from the perturbed data distribution. This process effectively undoes the degradation caused by the forward diffusion process. The result is a highly flexible and tractable generative model that can accurately model complex data distributions from random noise.

\subsection{Variational Perspective}

The Variational Perspective category includes models that use variational inference to approximate the target distribution, generally by minimizing the Kullback-Leibler divergence between the approximate and target distributions. Denoising Diffusion Probabilistic Models (DDPMs) \cite{sohl2015deep, ho2020denoising} are an example of this type of model, as they use a variational inference approach to estimate the parameters of a diffusion process.

\subsection{Denoising Diffusion Probabilistic Models (DDPMs)
}
\textbf{Forward Process.} DDPM defines the forward diffusion process as a Markov Chain where Gaussian noise is added in successive steps to obtain a set of noisy samples. Consider $q(x_{0})$ as the uncorrupted (original) data distribution. Given a data sample  $x_{0} \sim q(x_{0})$, a forward noising process $p$ which produces latent $x_{1}$ through $x_{T}$ by adding Gaussian noise at time t is defined as follows:
\begin{equation}
    q( x_ {t} |x_ {t-1} )=N( x_ {t} ; \sqrt {1-\beta _ {t}} \cdot x_ {t-1} , \beta _ {t} \cdot I), \forall t \in \{1,...,T\}
    \label{eq: 1}
\end{equation}

where $T$ and $\beta_{1},...,\beta_{T} \in [0,1)$ represent the number of diffusion
 steps and the variance schedule across diffusion steps, respectively.
 \textbf{I} is the identity matrix and $N(x;\mu,\sigma)$ represents the normal distribution of mean $\mu$ and covariance $\sigma$. Considering $\alpha_t = 1 - \beta_t$ and  $ \overline{\alpha}_t = \prod_{s=0}^{t} \alpha_s $, one can directly sample an arbitrary step of the noised
 latent conditioned on the input $x_{0}$ as follows:

\begin{equation}
    q(x_ {t} |x_ {0})=N(x_ {t} ; \sqrt {\overline{\alpha _ {t}}}  x_ {0}  ,(1-  \overline {\alpha }_ {t}  )I)
    \label{eq: 2}
\end{equation}
\begin{equation}
      x_ {t}  =  \sqrt {\overline {\alpha }_ {t}}  x_ {0}  +  \sqrt {1-\overline {\alpha }_ {l}\in}  
    \label{eq: 3}
\end{equation}

\begin{figure}
    \centering
    \includegraphics[width= \linewidth,]{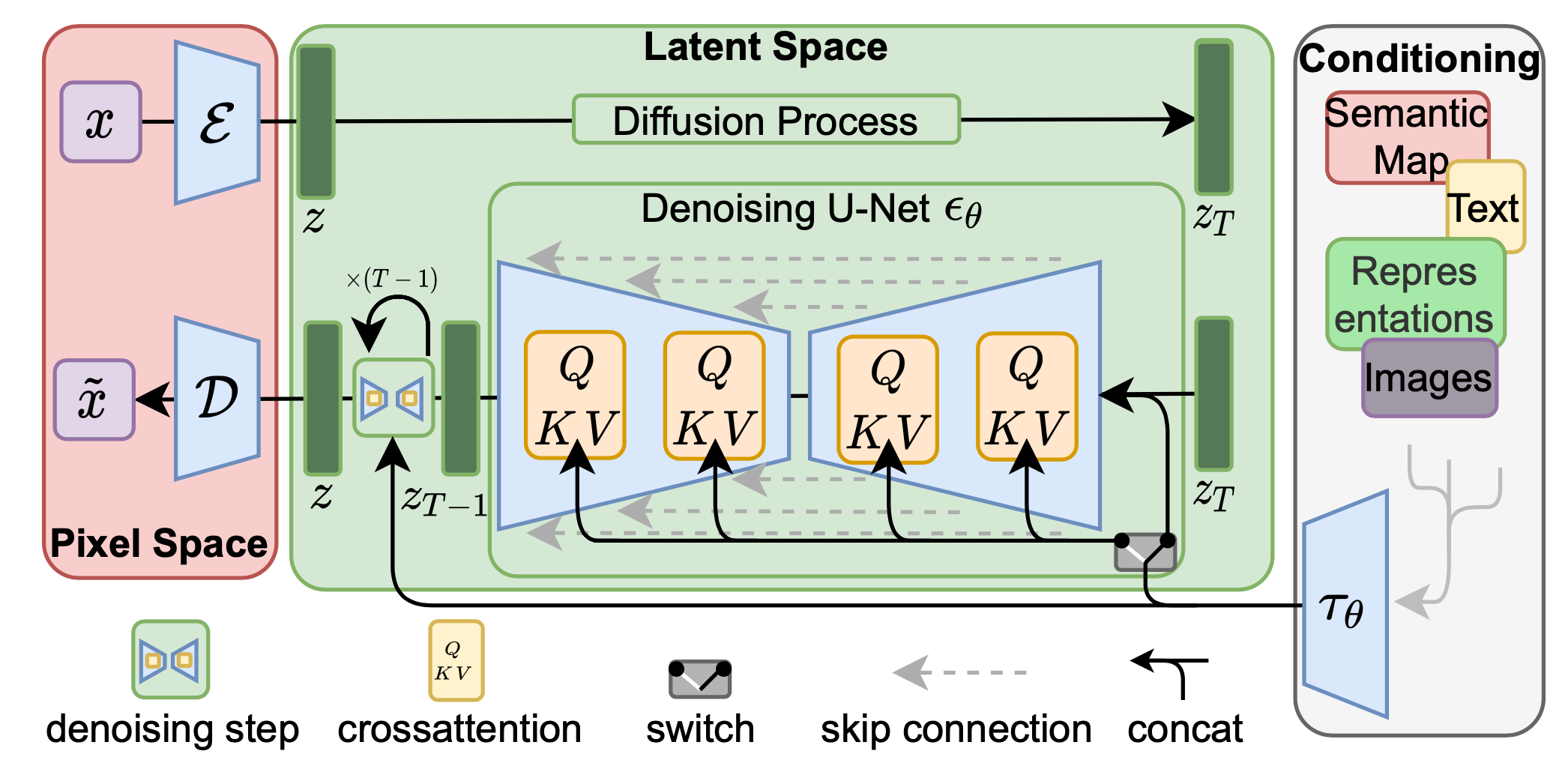}
    \caption{Architecture of Diffusion Model \cite{rink2021memory}}
     \label{fig: 3.3}
\end{figure}

\textbf{Reverse Process.} Leveraging the above definitions, we can approximate a reverse process to get a sample from $q(x_0)$. To this end, we can parameterize this reverse process by starting at $p(x_T)$ = $N (x_T;\textbf{0},\textbf{I})$ as follows:

\begin{equation}
    p_\theta\left(\mathbf{x}_{0: T}\right)=p\left(\mathbf{x}_T\right) \prod_{t=1}^T p_\theta\left(\mathbf{x}_{t-1} \mid \mathbf{x}_t\right)
    \label{eq: 4}
\end{equation}

\begin{equation}
    p_\theta\left(\mathbf{x}_{t-1} \mid \mathbf{x}_t\right)=N\left(\mathbf{x}_{t-1} ; \mu_\theta\left(\mathbf{x}_t, t\right), \Sigma_\theta\left(\mathbf{x}_t, t\right)\right) \text {. }
    \label{eq: 5}
\end{equation}

To train this model such that $p(x_0)$ learns the true data distribution $q(x_0)$, we can optimize the following variational bound on negative log-likelihood:

\begin{equation}
    \begin{aligned}
\mathbb{E}\left[-\log p_\theta\left(\mathbf{x}_0\right)\right] & \leq \mathbb{B}_q\left[-\log \frac{p_\theta\left(\mathbf{x}_{0: T}\right)}{q\left(\mathbf{x}_{1: T} \mid \mathbf{x}_0\right)}\right] \\
& =\mathbb{E}_q\left[-\log p\left(\mathbf{x}_T\right)-\sum_{t \geq 1} \log \frac{p_\theta\left(\mathbf{x}_{t-1} \mid \mathbf{x}_t\right)}{q\left(\mathbf{x}_t \mid \mathbf{x}_{t-1}\right)}\right] \\
& =-L_{\text {VL.B. }} .
\end{aligned}
\label{eq: 6}
\end{equation}

 Ho et al. \cite{ho2020denoising} found it better not to directly parameterize $\mu_\theta (x_t,t)$ as a neural network, but instead to train a model $\in_\theta (x_t,t)$
 to predict $\in$. Hence, by reparameterizing Equation (\ref{eq: 6}), they proposed a simplified objective as follows:
\begin{equation}
    L_{\text {simple }}=E_{t, x_0, \epsilon}\left[\left\|\epsilon-\epsilon_\theta\left(x_t, t\right)\right\|^2\right] \text {, }
    \label{eq: 7}
\end{equation}
 where the authors draw a connection between the loss in Equation (\ref{eq: 6}) to generative score networks in Song et al. \cite{song2019generative}.
\section{Proposed Methodology}
\begin{figure}[h]
    \centering
    \includegraphics[width= \linewidth
    ]{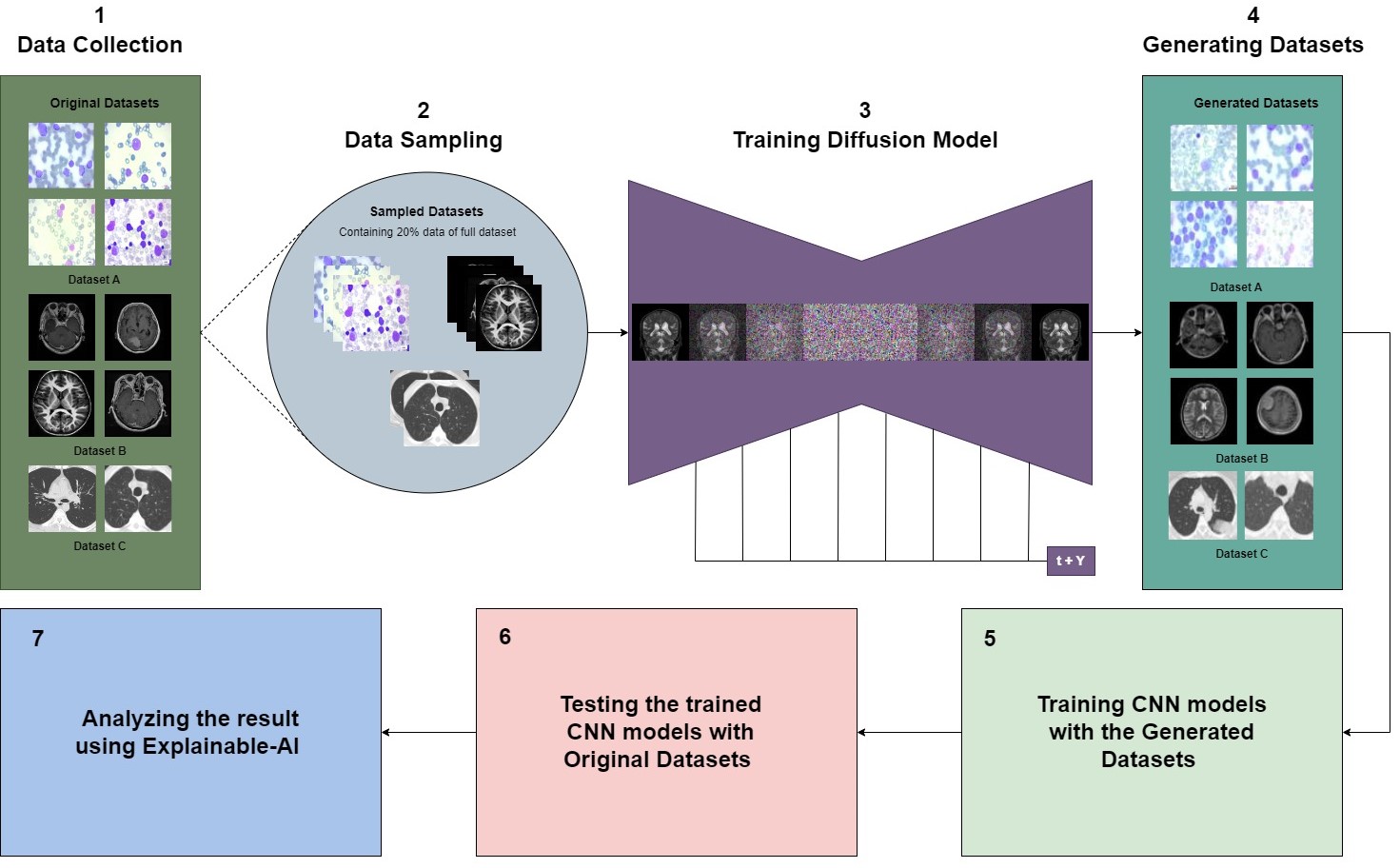}
    \caption{Flow diagram of our proposed methodology.}
     \label{fig: 4.1}
\end{figure}

This study investigates the potential of diffusion models (DMs) for generating synthetic medical images and their subsequent use in training Convolutional Neural Networks (CNNs) for various medical image analysis tasks. This section details the methodology employed, encompassing data acquisition, diffusion model training, synthetic data generation, CNN training, model evaluation, and explainable AI (XAI) analysis. The framework of our proposed methodology is illustrated in figure \ref{fig: 4.1}

\subsection{Data Collection}
\begin{figure}[h]
    \centering
    \includegraphics[width=\linewidth]{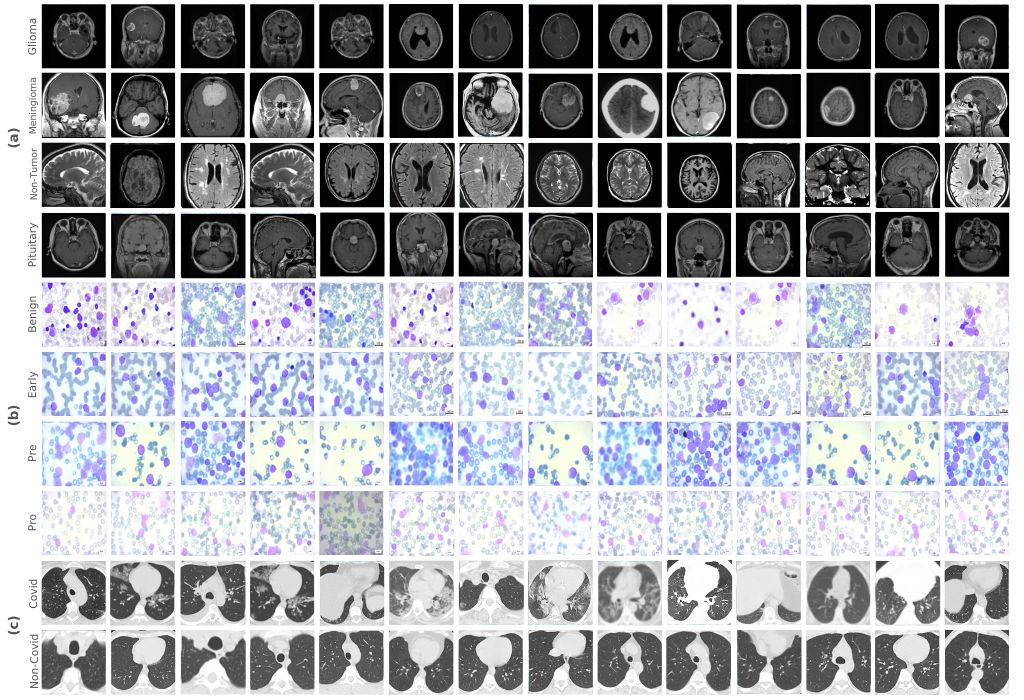}
    \caption{Sample images from the datasets a. Brain Tumor MRI b. Acute Lymphoblastic Leukemia (ALL) c. SARS-CoV-2 CT-Scans}
     \label{fig: 4.2}
\end{figure}

Three publicly available medical image datasets were obtained from Kaggle to represent a diverse range of medical imaging modalities and analysis tasks:
\begin{itemize}
    \item \textbf{Brain Tumor MRI:} This dataset consists of Magnetic Resonance Imaging (MRI) scans of patients diagnosed with and without brain tumors. MRI scans provide excellent soft tissue contrast, making them ideal for studying brain abnormalities.  Choosing this dataset allows us to explore the effectiveness of DM-generated synthetic data in tasks like tumor segmentation or anomaly detection in brain scans.

    \item \textbf{Acute Lymphoblastic Leukemia (ALL):} This dataset includes microscopic images of blood smears from patients with ALL, a type of blood cancer, and healthy controls. These images capture the morphology of blood cells, which is crucial for diagnosing various blood disorders.  This dataset helps evaluate the ability of DM-generated data to improve CNN performance in tasks like cell classification for blood analysis.

    \item \textbf{SARS-CoV-2 CT-Scans:} This dataset contains chest CT (Computed Tomography) scans from individuals with COVID-19 and healthy controls. CT scans provide detailed anatomical information about internal organs. Including this dataset allows us to assess the generalizability of the proposed approach to different medical imaging modalities beyond MRI and microscopic images. It also opens possibilities for exploring tasks like lung infection segmentation or anomaly detection in chest CT scans.
\end{itemize}
The chosen datasets represent a variety of medical imaging modalities (MRI, microscopy, CT scan) and address diverse medical conditions (brain tumors, blood cancer, respiratory illness). This selection aims to assess the generalizability of the proposed methodology (DM-generated data for CNN training) across different imaging techniques and disease contexts.
Despite the dataset diversity, it's important to acknowledge limitations. The findings might need further validation on a broader range of medical image datasets encompassing various organs, diseases, and imaging modalities. The number of images in each class is shown in Table \ref{table 4.1}. A sample image of the dataset is shown in the given Figure \ref{fig: 4.2}.

\begin{table}[]
\caption{Details of the datasets.}
\label{table 4.1}
\resizebox{\textwidth}{!}{%
\begin{tabular}{lclccc}
\toprule
\textbf{Dataset Name}                       & \textbf{Image Type}  & \textbf{Class Name} & \textbf{Total Images (Org)} & \textbf{Sampled Images} & \textbf{Generated Images} \\ \midrule
\multirow{4}{*}{\textbf{Brain Tumor MRI}} &  \multirow{4}{*}{\textbf{MRI}}         & Glioma     & 1621               & 325            & 1700             \\ \cmidrule(l){3-6} 
                                   &             & Meningioma & 1645               & 330            & 1700             \\ \cmidrule(l){3-6}
                                   &             & Notumor    & 2000               & 400            & 1700             \\ \cmidrule(l){3-6}
                                   &             & Pituitary  & 1757               & 350            & 1700             \\ \midrule
\multirow{4}{*}{\textbf{Acute Lymphoblastic Leukemia (ALL)}} & \multirow{4}{*}{\textbf{Microscopic}} & Benign     & 504                & 101            & 1000             \\ \cmidrule(l){3-6}
                                   &             & Early      & 985                & 197            & 1000             \\ \cmidrule(l){3-6}
                                   &             & Pre        & 963                & 193            & 1000             \\ \cmidrule(l){3-6}
                                   &             & Pro        & 804                & 161            & 1000             \\ \midrule
\multirow{2}{*}{\textbf{SARS-CoV-2 CT-Scans}}               & \multirow{2}{*}{\textbf{CT}}          & Covid      & 1252               & 251            & 1500             \\ \cmidrule(l){3-6}
                                   &             & Non-Covid  & 1229               & 246            & 1500             \\ \bottomrule
\end{tabular}
}
\end{table}
\subsection{Data Sampling}
To facilitate efficient training of the diffusion model, a stratified random sample (20\%) was extracted from each original dataset. This ensures class representation in the training data, maintaining the balance between different categories (e.g., tumor vs. healthy tissue in brain scans). 
\\[1\baselineskip]
While the original datasets might be vast, training a diffusion model effectively often requires a manageable amount of data. To address this, a stratified random sampling approach was employed. Stratification ensures that the training data maintains a proportional representation of each class present in the original dataset. For instance, the Brain Tumor MRI dataset might be stratified into "tumor present" and "tumor absent" classes. Following stratification, a random sample was drawn from each class, typically constituting 20\% of the total data points within that class. This sampling technique ensures the selected subset reflects the statistical distribution of the entire dataset while maintaining class balance. The number of sampled images can be seen from table \ref{table 4.1}
The rationale behind stratified random sampling is twofold:
\begin{itemize}
    \item \textbf{Efficiency:} Training a diffusion model on a smaller, representative subset of the data is computationally more efficient compared to using the entire dataset.
    \item \textbf{Class Imbalance Mitigation:} Medical image datasets can sometimes suffer from class imbalance, where one class (e.g., disease) is significantly underrepresented compared to others (e.g., healthy). Stratification ensures that the training data for the diffusion model includes a sufficient number of examples from each class, preventing the model from becoming biased toward the majority class.
    
\end{itemize}

\subsection{Training Diffusion Model}
This section details the training process for the diffusion model (DM) employed in this study. Diffusion models (DMs) are a class of generative models that learn to progressively remove noise added to real data points, ultimately enabling them to generate new, realistic samples. While this study does not utilize a conditional diffusion model (CDM) that incorporates additional information during noise removal, the chosen DM architecture leverages the core principles of noise diffusion effectively.
\\[1\baselineskip]
The specific DM architecture chosen for this study was [insert specific architecture reference here (e.g., U-Net-Based diffusion model)]. This architecture is well-suited for medical image data due to its ability to capture complex spatial relationships within the images.
The core components of the DM include:
\begin{itemize}
    \item \textbf{Forward Process.} DM defines the forward diffusion process as a Markov Chain where Gaussian noise is added in successive steps to obtain a set of noisy samples. Consider $q(x_{0})$ as the uncorrupted (original) data distribution. Given a data sample  $x_{0} \sim q(x_{0})$, a forward noising process $p$ which produces latent $x_{1}$ through $x_{T}$ by adding Gaussian noise at time t is defined as follows:
\begin{equation}
    q( x_ {t} |x_ {t-1} )=N( x_ {t} ; \sqrt {1-\beta _ {t}} \cdot x_ {t-1} , \beta _ {t} \cdot I), \forall t \in \{1,...,T\}
    \label{eq: 1}
\end{equation}

where $T$ and $\beta_{1},...,\beta_{T} \in [0,1)$ represent the number of diffusion
 steps and the variance schedule across diffusion steps, respectively.
 \textbf{I} is the identity matrix and $N(x;\mu,\sigma)$ represents the normal distribution of mean $\mu$ and covariance $\sigma$. Considering $\alpha_t = 1 - \beta_t$ and  $ \overline{\alpha}_t = \prod_{s=0}^{t} \alpha_s $, one can directly sample an arbitrary step of the noised
 latent conditioned on the input $x_{0}$ as follows:

\begin{equation}
    q(x_ {t} |x_ {0})=N(x_ {t} ; \sqrt {\overline{\alpha _ {t}}}  x_ {0}  ,(1-  \overline {\alpha }_ {t}  )I)
    \label{eq: 2}
\end{equation}
\begin{equation}
      x_ {t}  =  \sqrt {\overline {\alpha }_ {t}}  x_ {0}  +  \sqrt {1-\overline {\alpha }_ {l}\in}  
    \label{eq: 3}
\end{equation}

\begin{equation}
    N = [(I_w - P_w) / S + 1] * [(I_h - P_h) / S + 1]
    \label{eq 1}
\end{equation}

\item \textbf{Reverse Process.} Leveraging the above definitions, we can approximate a reverse process to get a sample from $q(x_0)$. To this end, we can parameterize this reverse process by starting at $p(x_T)$ = $N (x_T;\textbf{0},\textbf{I})$ as follows:

\begin{equation}
    p_\theta\left(\mathbf{x}_{0: T}\right)=p\left(\mathbf{x}_T\right) \prod_{t=1}^T p_\theta\left(\mathbf{x}_{t-1} \mid \mathbf{x}_t\right)
    \label{eq: 4}
\end{equation}

\begin{equation}
    p_\theta\left(\mathbf{x}_{t-1} \mid \mathbf{x}_t\right)=N\left(\mathbf{x}_{t-1} ; \mu_\theta\left(\mathbf{x}_t, t\right), \Sigma_\theta\left(\mathbf{x}_t, t\right)\right) \text {. }
    \label{eq: 5}
\end{equation}

To train this model such that $p(x_0)$ learns the true data distribution $q(x_0)$, we can optimize the following variational bound on negative log-likelihood:

\begin{equation}
    \begin{aligned}
\mathbb{E}\left[-\log p_\theta\left(\mathbf{x}_0\right)\right] & \leq \mathbb{B}_q\left[-\log \frac{p_\theta\left(\mathbf{x}_{0: T}\right)}{q\left(\mathbf{x}_{1: T} \mid \mathbf{x}_0\right)}\right] \\
& =\mathbb{E}_q\left[-\log p\left(\mathbf{x}_T\right)-\sum_{t \geq 1} \log \frac{p_\theta\left(\mathbf{x}_{t-1} \mid \mathbf{x}_t\right)}{q\left(\mathbf{x}_t \mid \mathbf{x}_{t-1}\right)}\right] \\
& =-L_{\text {VL.B. }} .
\end{aligned}
\label{eq: 6}
\end{equation}
\end{itemize}

This section detailed the methodology employed for training the diffusion model. The chosen architecture leverages the information from labeled data (images and corresponding labels) to enhance the quality and relevance of the generated synthetic medical images. The training process involves iteratively adding noise to real medical images and their labels, and then training the model to predict and remove this noise effectively while considering the label information. This approach allows the DM to potentially generate synthetic images that share the characteristics and statistical properties of the original data while being conditioned on the provided labels. Future work could explore the impact of different label conditioning strategies and their influence on the quality and targeted nature of the generated synthetic medical images.
\subsection{Generating Datasets}
Once the diffusion model (DM) has been trained effectively, it can be leveraged to generate a significantly larger dataset of synthetic medical images. This expanded dataset addresses potential data scarcity challenges often encountered in various medical image analysis tasks. This section details the process of generating synthetic data using the trained DM.
\\[1\baselineskip]
The core functionality of the trained DM lies in its ability to progressively remove noise added to real medical images. This capability is harnessed for synthetic data generation by reversing the noise addition process implemented during training. Here's a breakdown of the steps involved:

\begin{enumerate}
    \item \textbf{Initial Noise Injection:} A random noise vector is sampled from the same noise distribution used during training (e.g., Gaussian noise).

    \item \textbf{Iterative Noise Removal:} The trained DM takes the initial noise vector and a random timestep $t$ as input. It then iteratively predicts the noise to be removed at $t-1$ each timestep, effectively reversing the noise addition process from training.

    \item \textbf{Image Reconstruction:} The predicted noise across all timesteps is progressively removed from the initial noise vector to obtain a reconstructed image. This reconstructed image represents a newly generated synthetic medical image.

    \item \textbf{Iteration:} Steps 2 and 3 can be repeated multiple times to generate a large number of synthetic medical images. The DM, having learned the underlying distribution of the real data during training, can generate new images that statistically resemble the original dataset while potentially inheriting the label information if the DM architecture incorporates label conditioning.

\end{enumerate}
\subsection{Training CNN Models}
This section details the training process for Convolutional Neural Networks (CNNs) employed for various medical image analysis tasks. The synthetic data generated using the diffusion model serves as the primary training data for these CNNs.Eight distinct CNN architectures were designed and implemented to address the specific medical image analysis tasks associated with the chosen datasets (e.g., tumor segmentation, cell classification, anomaly detection). Exploring a diverse range of CNN architectures allows us to identify models that are well-suited for the specific characteristics of the medical image analysis tasks and the generated synthetic data. This approach mitigates the risk of overfitting to a single architecture and potentially leads to the selection of a more robust and generalizable CNN model.

Steps followed in the training process:
\begin{itemize}
    \item \textbf{Data Preprocessing:} The generated synthetic medical images were preprocessed for CNN training. This involves techniques like image normalization, resizing to a standard size, or data augmentation to improve model performance and generalization.

    \item \textbf{Model Training:} Each CNN architecture was trained independently using the preprocessed synthetic medical images and their corresponding labels. The AdamW optimizer was employed for efficient gradient descent during backpropagation. The cross-entropy loss function was used to measure the discrepancy between the predicted labels (model output) and the true labels associated with the synthetic images.

    \item \textbf{5-Fold Cross-Validation:} To mitigate overfitting and assess model generalizability, a 5-fold cross-validation approach was implemented. The synthetic data was split into five folds. In each fold, four folds were used for training, and the remaining fold was used for validation. This process was repeated five times, ensuring that each data point was used for validation once. The final model performance was evaluated based on the average performance across all five folds.

    \item \textbf{Hyperparameter Tuning:} Hyperparameter tuning techniques were employed to optimize the performance of each CNN architecture. This involves adjusting learning rates, batch sizes, or other hyperparameters to achieve the best possible performance on the validation data.

    \item \textbf{Early Stopping:} Early stopping was implemented to prevent overfitting. Training was terminated if the validation loss did not improve for a predefined number of epochs.

    \item \textbf{Epochs:} The training process for each CNN architecture was conducted for a maximum of 50 epochs. Training might be stopped earlier based on the early stopping criteria.

\end{itemize}
\subsection{Model Evaluation}
The performance of each trained CNN model was evaluated using a hold-out validation approach. The remaining 80\% (untouched) portion of the original datasets was used for testing. This approach aimed to assess the effectiveness of the generated synthetic data in training robust and generalizable CNN models for medical image analysis tasks. Performance metrics appropriate for the specific medical image analysis tasks (e.g., accuracy, precision, recall, F1-score) are used to evaluate the CNN models.

\textbf{Precision :} Precision serves as an evaluation measure in tasks involving classification, like machine learning or information retrieval. It is employed to gauge the correctness of positive predictions generated by a model. Precision calculates the proportion of true positive predictions among all the positive predictions made by the model. 
\begin{equation}
    Precision = \frac{TP}{TP+FP}
\end{equation}
\textbf{Recall: }Recall, which can also be called Sensitivity or True Positive Rate, is a performance metric utilized in classification tasks, like those in machine learning. Its purpose is to evaluate how well a model can accurately recognize all the positive instances within the dataset. In essence, it measures the proportion of true positive predictions relative to the overall actual positives.
\begin{equation}
    Recall = \frac{TP}{TP+FN}
\end{equation}
\textbf{F1-score: }The F1 Score is a performance measure applied in classification tasks, including machine learning and data analysis, to establish a trade-off between precision and recall. It proves valuable when there is a requirement to strike a balance between false positives (precision) and false negatives (recall) and achieve a middle ground between the two. 
\begin{equation}
    F1-score = 2 \times \frac{precision \times recall}{precision + recall} 
\end{equation}
\textbf{Accuracy: }Accuracy refers to the ratio of the total number of correct predictions to the total number of input samples. It simply measures the percentage of correct predictions that a machine-learning model has made.
\begin{equation}
   Accuracy = \frac{TP + TN}{TP+TN+FP+FN} 
\end{equation}
Where, 
\begin{itemize}
    \item 	\textbf{True Positive (TP):}  A sample is predicted to be positive and its label is actually positive.
    \item \textbf{	True Negative (TN):}  A sample is predicted to be negative and its label is actually negative.
    \item \textbf{False Positive (FP):} A sample is predicted to be positive and its label is actually negative.
    \item\textbf{	False Negative (FN):} A sample is predicted to be negative and its label is actually positive.
\end{itemize}
\subsection{Explainable AI (XAI) Analysis}
Understanding the rationale behind a CNN's predictions, particularly in medical image analysis tasks, is crucial for building trust and ensuring the interpretability of the model's decisions. This section details the application of Local Interpretable Model-agnostic Explanations (LIME) to analyze the inner workings of the trained CNN models.
\\[1\baselineskip]
LIME is a technique that allows us to explain individual predictions made by a complex model, such as a CNN. It works by approximating the local behavior of the model around a specific prediction using an interpretable model, such as a linear regression model. This approach provides insights into the features in the input image that contribute most significantly to the model's prediction.
\\[1\baselineskip]
In this study, LIME was employed to explain the predictions of the CNN models on a subset of the synthetic medical images. Here's how the analysis was conducted:

\begin{itemize}
    \item \textbf{Image Selection:} A representative set of synthetic medical images was chosen for LIME analysis. This selection might include images where the CNN model's prediction is correct, incorrect, or exhibits a high degree of uncertainty.

    \item \textbf{Image Selection:} For each selected image, LIME was used to generate an explanation for the CNN's prediction. This explanation takes the form of a heatmap highlighting the image regions that the model relied on most heavily for its decision. Additionally, LIME might provide a list of the most influential features (e.g., pixel intensities, edges, textures) contributing to the prediction.

    \item \textbf{Image Selection:} The generated LIME explanations were analyzed to understand how the CNN model utilizes the features within the synthetic images to arrive at its predictions. This analysis can reveal potential biases or limitations in the model's decision-making process and inform strategies for further improvement.

\end{itemize}
By leveraging LIME, this study aims to gain valuable insights into the interpretability of the CNN models trained on synthetic data. This understanding can contribute to building trust in the models' predictions and ultimately improve their effectiveness in real-world medical image analysis tasks.

\section{Results and Discussion}
\subsection{Experiment Setup}
The experiment is conducted on Google Colab and Jupyter Notebook in a system with an Intel(R) Core(TM) i5-7200U CPU processor and 8 GB RAM.
An A100 8 GB GPU is used for training the diffusion model.

\subsection{Results}
This section presents the results of our investigation into the effectiveness of diffusion models for generating synthetic medical images for training CNN models in medical image analysis tasks. We focus on three datasets: SARS-CoV-2 CT-Scans, brain tumor MRI scans, and Acute Lymphoblastic Leukemia (ALL).

\subsection{Diffusion Model Training and Synthetic Data Generation}
\begin{figure}[h]
    \centering
    \includegraphics[width=\linewidth]{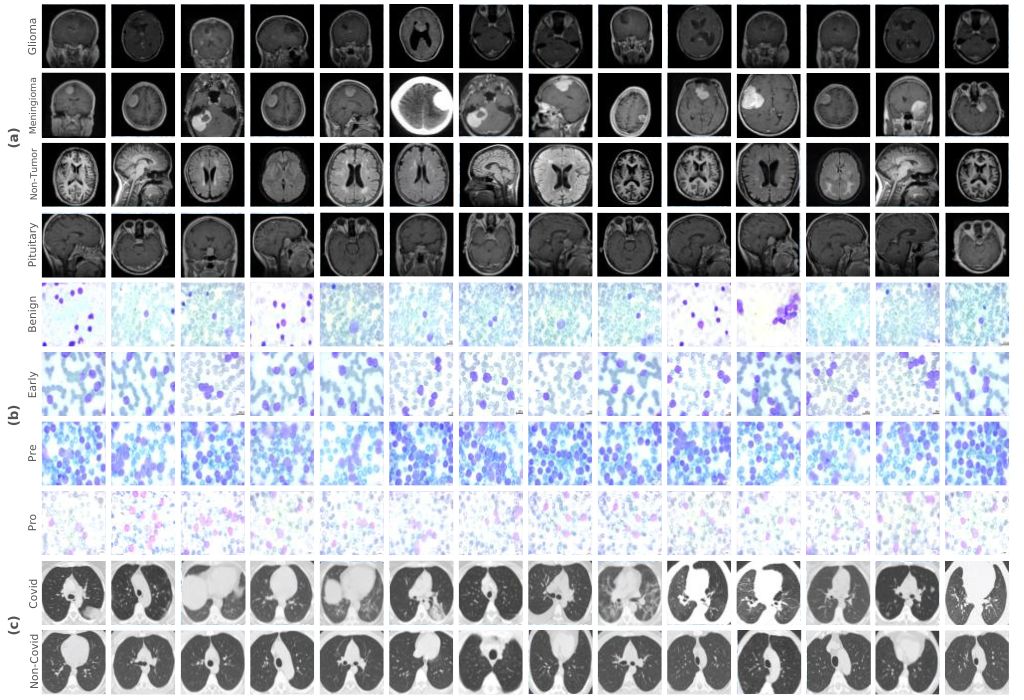}
    \caption{Sample images from the generated datasets \textbf{(a)} Brain Tumor MRI \textbf{(b)} Acute Lymphoblastic Leukemia (ALL) \textbf{(c)} SARS-CoV-2 CT-Scans}
     \label{fig: 5.1}
\end{figure}

The diffusion model was trained on an 8GB A100 GPU for 4 days. Following successful training, the model generated a new synthetic dataset for each of the three medical domains. Some sample images from the generated datasets are shown in figure \ref{fig: 5.1}. The size and characteristics of these synthetic datasets will be further explored in future work.

\subsection{Evaluation of CNN Models on Synthetic and Real Data}
Eight pre-trained CNN architectures were evaluated on the classification tasks: ResNet-19, ResNet-50, VGG-16, VGG-19, AlexNet, DenseNet-121, MobileNetV2, and GoogleNet. Each model was trained on the generated synthetic dataset for each medical domain and subsequently evaluated on an unseen 80\% split of the original real data. Test accuracy, precision, recall, and F1-score were calculated for each model-dataset combination. The experimental results are shown in table \ref{table 5.1}.
\begin{itemize}
    \item \textbf{SARS-CoV-2 CT-Scans:} Among the eight CNN architectures evaluated on the Covid-19 dataset, ResNet-50 achieved the highest performance on the unseen test set. The model trained on synthetic data achieved a test accuracy of 78.24\%. Precision, recall, and F1-score were all found to be 77\%. These results suggest that the diffusion model was able to generate synthetic Covid-19 chest X-rays that captured essential features for accurate classification. However, further investigation is needed to understand if this performance can be improved through hyperparameter tuning of the diffusion model or the training process of the CNN model.

    \item \textbf{Brain Tumor MRI:} VGG-19 achieved the best performance on the brain tumor dataset, with a test accuracy of 86.46\% on the unseen data split. Additionally, the model achieved a precision of 89\%, a recall of 87\%, and an F1-score of 87\%. These results indicate that the synthetic brain tumor MRI scans generated by the diffusion model were highly informative for the classification task. VGG-19's superior performance in this domain could be due to its deeper architecture compared to other models, potentially allowing it to capture more complex features within the brain tumor images. 

    \item \textbf{Acute Lymphoblastic Leukemia (ALL):} DenseNet-121 emerged as the top performer on the Leukemia dataset. The model trained on synthetic data achieved a test accuracy of 91.38\%. Precision, recall, and F1-score were all found to be 92\%. This finding suggests that the diffusion model was particularly effective in generating synthetic Leukemia blood smear images that closely resembled real-world data. DenseNet-121's dense connectivity architecture might have been advantageous in capturing the subtle variations within these images. 
    
\end{itemize}
\subsection{ LIME Explainability Analysis}
\begin{figure}[h]
    \centering
    \includegraphics[width=\linewidth]{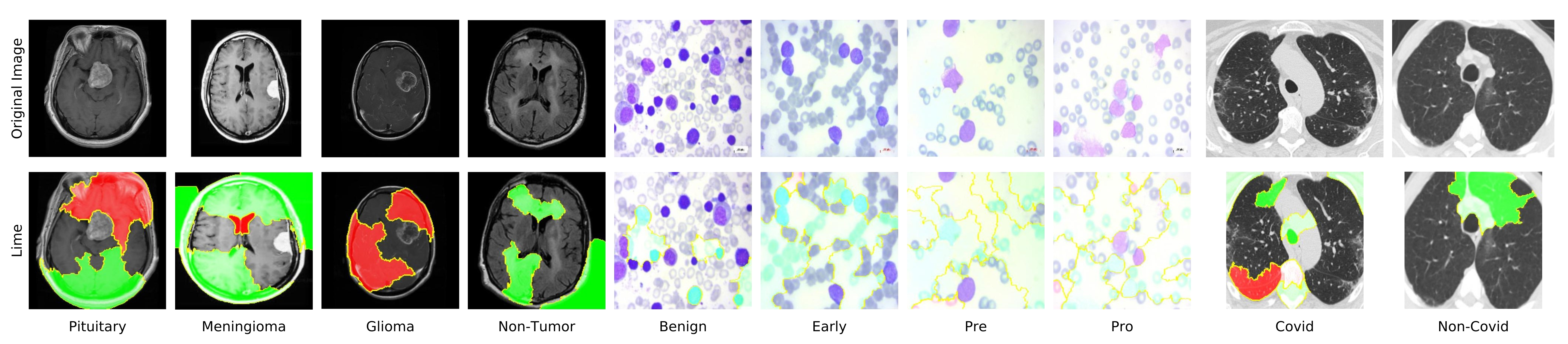}
    \caption{Local Interpretable Model-Agnostic Explanations (LIME)}
     \label{fig: 5.3}
\end{figure}

To gain deeper insights into the decision-making process of the top-performing models on each dataset (ResNet-50 for Covid-19, VGG-19 for Brain Tumor, and DenseNet-121 for Leukemia), we employed Local Interpretable Model-Agnostic Explanations (LIME). LIME is a technique that allows us to explain the predictions of any black-box model by approximating it locally with an interpretable model around a specific prediction. This helps us understand which features within the medical images were most influential for the model's classification. The results for the LIME are shown in figure \ref{fig: 5.3} For each dataset, we applied LIME to analyze several image classifications from the unseen test set.

\begin{table}[]
\caption{Comparison of CNN model performance on unseen real data after training on synthetic medical images}
\label{table 5.1}
\resizebox{\textwidth}{!}{%
\begin{tabular}{llccccccccc}
\toprule
\textbf{Model Name}    & \textbf{Dataset}     & \textbf{Train loss} & \textbf{Val loss} & \textbf{Test loss} & \textbf{ Train Acc} & \textbf{Val Acc} & \textbf{Test Acc} & \textbf{Precision} & \textbf{Recall} & \textbf{F1-Score}   \\ \midrule
\textbf{Resnet18}      & Covid       & 0.0103     & 0.0232   & 2.2450    & 99.82     & 99.55   & 76.33    & 0.75      & 0.75   & 0.75 \\  \cmidrule(l){2-11}
              & Brain Tumor & 0.0179     & 0.0372   & 1.2192    & 99.68     & 98.99   & 82.89    & 0.87      & 0.84   & 0.84 \\ \cmidrule(l){2-11}
              & ALL         & 0.0287     & 0.0149   & 1.1923    & 99.86     & 99.67   & 86.94    & 0.87      & 0.86   & 0.86 \\ \midrule
\textbf{ResNet50}      & Covid       & 0.0083     & 0.0320   & 1.8761    & 99.86     & 99.55   & \textbf{78.24}    & \textbf{0.77}      & \textbf{0.77}   & \textbf{0.77} \\ \cmidrule(l){2-11}
              & Brain Tumor & 0.0074     & 0.0734   & 1.0452    & 99.79     & 98.80   & 85.15    & \textbf{0.89}      & \textbf{0.87}   & \textbf{0.87} \\ \cmidrule(l){2-11}
              & ALL         & 0.0166     & 0.0165   & 0.8619    & 99.95     & 99.51   & 88.12    & 0.89      & 0.88   & 0.88 \\ \midrule
\textbf{VGG16}         & Covid       & 0.0074     & 0.0154   & 0.7655    & 99.84     & 99.51   & 77.67    & 0.77      & 0.74   & 0.74 \\ \cmidrule(l){2-11}
              & Brain Tumor & 0.0077     & 0.0399   & 0.7887    & 99.85     & 99.10   & 85.68    & 0.89      & 0.82   & 0.82 \\ \cmidrule(l){2-11}
              & ALL         & 0.0078     & 0.0136   & 1.3405    & 99.90     & 99.70   & 83.01    & 0.88      & 0.79   & 0.78 \\ \midrule
\textbf{VGG19}         & Covid       & 0.0079     & 0.0158   & 0.7703    & 99.89     & 99.60   & 75.73    & 0.75      & 0.74   & 0.73 \\ \cmidrule(l){2-11}
              & Brain Tumor & 0.0100     & 0.0589   & 0.7354    & 99.81     & 99.07   & \textbf{86.46}    & 0.88      & 0.84   & 0.85 \\ \cmidrule(l){2-11}
              & ALL         & 0.0063     & 0.0105   & 1.2673    & 99.91     & 99.67   & 82.38    & 0.86      & 0.77   & 0.76 \\ \midrule
\textbf{AlexNet}       & Covid       & 0.0369     & 0.0383   & 0.9169    & 98.97     & 98.70   & 72.18    & 0.71      & 0.70   & 0.70 \\ \cmidrule(l){2-11}
              & Brain Tumor & 0.0360     & 0.0875   & 0.9130    & 99.20     & 98.37   & 79.86    & 0.85      & 0.75   & 0.76 \\ \cmidrule(l){2-11}
              & ALL         & 0.0236     & 0.0318   & 0.9822    & 99.49     & 99.21   & 82.22    & 0.86      & 0.78   & 0.77 \\ \midrule
\textbf{DenseNet121}   & Covid       & 0.0132     & 0.0172   & 1.8653    & 99.71     & 99.33   & 76.36    & 0.76      & 0.76   & 0.75 \\ \cmidrule(l){2-11}
              & Brain Tumor & 0.0059     & 0.0521   & 0.9798    & 99.84     & 98.99   & 85.09    & 0.89      & 0.86   & 0.87 \\ \cmidrule(l){2-11}
              & ALL         & 0.0286     & 0.0118   & 0.6045    & 99.94     & 99.73   & \textbf{91.38}    & \textbf{0.92}      & \textbf{}{0.91}   & \textbf{0.92} \\ \midrule
\textbf{MobileNet\_v2} & Covid       & 0.0205     & 0.0305   & 2.3197    & 99.36     & 98.92   & 75.98    & 0.75      & 0.75   & 0.75 \\ \cmidrule(l){2-11}
              & Brain Tumor & 0.0190     & 0.0856   & 1.1486    & 99.57     & 98.56   & 84.05    & 0.89      & 0.86   & 0.87 \\ \cmidrule(l){2-11}
              & ALL         & 0.0470     & 0.0264   & 1.0470    & 99.48     & 99.18   & 85.21    & 0.89      & 0.88   & 0.88 \\ \midrule
\textbf{GoogleNet}     & Covid       & 0.0165     & 0.0229   & 1.7411    & 99.71     & 99.33   & 75.83    & 0.75      & 0.75   & 0.75 \\ \cmidrule(l){2-11}
              & Brain Tumor & 0.0143     & 0.0475   & 0.9578    & 99.61     & 98.91   & 83.82    & 0.88      & 0.86   & 0.87 \\ \cmidrule(l){2-11}
              & ALL         & 0.0400     & 0.0254   & 0.5807    & 99.75     & 99.46   & 89.40    & 0.91      & 0.90   & 0.90 \\ \bottomrule
\end{tabular}
}
\end{table}
\subsection{Discussion}
This study investigated the potential of diffusion models for generating synthetic medical images to train CNN models for medical image analysis tasks. We evaluated the effectiveness of this approach on three datasets: Covid-19 chest X-rays, brain tumor MRI scans, and Leukemia blood smear images.
\\[1\baselineskip]
The results are encouraging, with all three datasets achieving promising classification performance using models trained on synthetic data. Notably, the Leukemia dataset reached a remarkable test accuracy of 91.38\%, suggesting diffusion models might be particularly adept at generating synthetic images for specific medical domains with well-defined visual characteristics. However, a deeper discussion is necessary to fully understand the implications and limitations of this approach.
\\[1\baselineskip]
The study demonstrates that diffusion models can generate synthetic medical images that can be effectively utilized for training CNN models. The top-performing models on each dataset achieved good test accuracy, precision, recall, and F1-score. While further optimization is possible, these results suggest that diffusion models have the potential to address data scarcity and bias limitations commonly encountered in medical image analysis.
\\[1\baselineskip]
However, some questions remain:
\begin{itemize}
    \item \textbf{Impact of Dataset Size and Diversity:} The size and diversity of the generated synthetic datasets were not extensively explored. Future work should investigate how these factors influence the performance of CNN models. Training models on synthetic datasets of varying sizes and employing data augmentation techniques during diffusion model training could provide valuable insights.

    \item \textbf{Generalizability of Synthetic Data:}  While the models performed well on unseen test splits within each dataset, further validation is needed. The generalizability of the approach should be assessed on entirely new datasets from the same medical domains.

    \item \textbf{Comparison with Traditional Techniques:}  It would be valuable to compare the performance of models trained on synthetic data with models trained on traditional data augmentation techniques or with limited real data. This comparison would provide a clearer picture of the relative advantages and limitations of using diffusion models.
\end{itemize}
Both the diffusion model and the CNN models could potentially benefit from hyperparameter tuning. Optimizing hyperparameters like the noise schedule, diffusion steps, learning rate, and optimizer choice could lead to further performance improvements. Techniques like grid search or Bayesian optimization can be employed to identify the optimal configurations for each model and dataset.
\\[1\baselineskip]
The preliminary LIME analysis suggests the top-performing models focused on relevant features for classification. However, a more in-depth exploration is crucial. Quantifying feature importance scores from LIME would reveal which image regions are most critical for the models' performance. Additionally, comparing LIME explanations for real and synthetic data would be insightful. If the models utilize comparable features for classification across both data types, it would strengthen the argument that the synthetic data effectively captures essential characteristics required for accurate image analysis.

\section{Conclusion and Future Work}
In conclusion, This study investigated the effectiveness of diffusion models for generating synthetic medical images for training CNN models in medical image analysis tasks. We focused on three datasets: SARS-CoV-2 CT-Scans, brain tumor MRI, and Acute Lymphoblastic Leukemia (ALL). The diffusion model was trained on the three datasets, successfully generating new synthetic datasets for each domain. Subsequently, eight pre-trained CNN architectures were evaluated on these synthetic datasets before being tested on unseen real data.
\\[1\baselineskip]
The results are encouraging. All three datasets achieved promising classification performance using models trained on synthetic data. Notably, the Leukemia dataset reached a test accuracy of 91.38\%, suggesting that diffusion models can be particularly effective for specific medical domains with well-defined visual characteristics. These findings demonstrate the potential of diffusion models to address the challenges of data scarcity and bias commonly encountered in medical image analysis.
\\[1\baselineskip]
However, this study also highlights the need for further exploration in several key areas. A more comprehensive analysis of the generated synthetic datasets is crucial. This includes investigating the impact of dataset size, diversity, and the distribution of features compared to real data. Techniques like training models on datasets of varying sizes, employing data augmentation during diffusion model training, and using statistical tests for distribution comparison can provide valuable insights.
\\[1\baselineskip]
Optimizing hyperparameters for both the diffusion model and the CNN models could lead to further performance improvements. Here, techniques like grid search or Bayesian optimization can be employed to identify the optimal configuration for the diffusion model's hyperparameters, while cross-validation can be used to fine-tune the hyperparameters of the CNN models trained on synthetic data.
\\[1\baselineskip]
To assess the generalizability of the approach, further validation is necessary. Evaluating the effectiveness of the diffusion model and the trained CNN models on entirely new datasets from the same medical domains, along with a performance comparison with models trained on traditional data augmentation techniques or with limited real data, would provide a clearer picture of the relative advantages and limitations of using diffusion models.
\\[1\baselineskip]
The preliminary LIME analysis provided valuable insights into the models' focus on relevant features for classification. However, a more in-depth exploration is warranted. Quantifying the feature importance scores generated by LIME would reveal which image regions are most critical for the models' performance. Additionally, a detailed comparison of LIME explanations for images from both real and synthetic datasets would be insightful. If the models utilize comparable features for classification across both data types, it would strengthen the argument that the synthetic data effectively captures the essential characteristics required for accurate image analysis.
\\[1\baselineskip]
By addressing these future work directions, we can gain a more comprehensive understanding of the effectiveness of diffusion models for generating synthetic medical images. This will contribute to the development of more robust and generalizable CNN models for various medical image analysis tasks, ultimately aiding in improved medical diagnosis and treatment planning.

\section*{Conflict of interest}
The authors declare no conflict of interest.
\section*{Use of Generative AI and AI-assisted Technologies}
During the preparation of this work the author(s) used ChatGPT in order to reduce grammatical errors and writing clarity. After using this tool/service, the author(s) reviewed and edited the content as needed and take(s) full responsibility for the content of the published article.
\bibliographystyle{unsrt}  
\bibliography{main}

\begin{thebibliography}{10}

\bibitem{litjens2017survey}
Geert Litjens, Thijs Kooi, Babak~Ehteshami Bejnordi, Arnaud Arindra~Adiyoso Setio, Francesco Ciompi, Mohsen Ghafoorian, Jeroen~Awm Van Der~Laak, Bram Van~Ginneken, and Clara~I S{\'a}nchez.
\newblock A survey on deep learning in medical image analysis.
\newblock {\em Medical image analysis}, 42:60--88, 2017.

\bibitem{yu2016automated}
Lequan Yu, Hao Chen, Qi~Dou, Jing Qin, and Pheng-Ann Heng.
\newblock Automated melanoma recognition in dermoscopy images via very deep residual networks.
\newblock {\em IEEE transactions on medical imaging}, 36(4):994--1004, 2016.

\bibitem{avvakumov2021envy}
Sergey Avvakumov and Roman Karasev.
\newblock Envy-free division using mapping degree.
\newblock {\em Mathematika}, 67(1):36--53, 2021.

\bibitem{pinaya2022brain}
Walter~HL Pinaya, Petru-Daniel Tudosiu, Jessica Dafflon, Pedro~F Da~Costa, Virginia Fernandez, Parashkev Nachev, Sebastien Ourselin, and M~Jorge Cardoso.
\newblock Brain imaging generation with latent diffusion models.
\newblock In {\em MICCAI Workshop on Deep Generative Models}, pages 117--126. Springer, 2022.

\bibitem{ozbey2023unsupervised}
Muzaffer {\"O}zbey, Onat Dalmaz, Salman~UH Dar, Hasan~A Bedel, {\c{S}}aban {\"O}zturk, Alper G{\"u}ng{\"o}r, and Tolga {\c{C}}ukur.
\newblock Unsupervised medical image translation with adversarial diffusion models.
\newblock {\em IEEE Transactions on Medical Imaging}, 2023.

\bibitem{chung2022score}
Hyungjin Chung and Jong~Chul Ye.
\newblock Score-based diffusion models for accelerated mri.
\newblock {\em Medical image analysis}, 80:102479, 2022.

\bibitem{wolleb2022diffusion}
Julia Wolleb, Robin Sandk{\"u}hler, Florentin Bieder, Philippe Valmaggia, and Philippe~C Cattin.
\newblock Diffusion models for implicit image segmentation ensembles.
\newblock In {\em International Conference on Medical Imaging with Deep Learning}, pages 1336--1348. PMLR, 2022.

\bibitem{zamzmi2020unified}
Ghada Zamzmi, Sivaramakrishnan Rajaraman, and Sameer Antani.
\newblock Unified representation learning for efficient medical image analysis.
\newblock {\em arXiv preprint arXiv:2006.11223}, 2020.

\bibitem{goodfellow2014generative}
Ian Goodfellow, Jean Pouget-Abadie, Mehdi Mirza, Bing Xu, David Warde-Farley, Sherjil Ozair, Aaron Courville, and Yoshua Bengio.
\newblock Generative adversarial nets.
\newblock {\em Advances in neural information processing systems}, 27, 2014.

\bibitem{isola2017image}
Phillip Isola, Jun-Yan Zhu, Tinghui Zhou, and Alexei~A Efros.
\newblock Image-to-image translation with conditional adversarial networks.
\newblock In {\em Proceedings of the IEEE conference on computer vision and pattern recognition}, pages 1125--1134, 2017.

\bibitem{tang2019ct}
You-Bao Tang, Sooyoun Oh, Yu-Xing Tang, Jing Xiao, and Ronald~M Summers.
\newblock Ct-realistic data augmentation using generative adversarial network for robust lymph node segmentation.
\newblock In {\em Medical Imaging 2019: Computer-Aided Diagnosis}, volume 10950, pages 976--981. SPIE, 2019.

\bibitem{popescu2021retinal}
Dan Popescu, Mihaela Deaconu, Loretta Ichim, and Grigore Stamatescu.
\newblock Retinal blood vessel segmentation using pix2pix gan.
\newblock In {\em 2021 29th Mediterranean Conference on Control and Automation (MED)}, pages 1173--1178. IEEE, 2021.

\bibitem{aljohani2022generating}
Abeer Aljohani and Nawaf Alharbe.
\newblock Generating synthetic images for healthcare with novel deep pix2pix gan.
\newblock {\em Electronics}, 11(21):3470, 2022.

\bibitem{sun2022pix2pix}
Jingzhang Sun, Yu~Du, ChienYing Li, Tung-Hsin Wu, BangHung Yang, and Greta~SP Mok.
\newblock Pix2pix generative adversarial network for low dose myocardial perfusion spect denoising.
\newblock {\em Quantitative imaging in medicine and surgery}, 12(7):3539, 2022.

\bibitem{zhu2017unpaired}
Jun-Yan Zhu, Taesung Park, Phillip Isola, and Alexei~A Efros.
\newblock Unpaired image-to-image translation using cycle-consistent adversarial networks.
\newblock In {\em Proceedings of the IEEE international conference on computer vision}, pages 2223--2232, 2017.

\bibitem{du2018reduction}
Muge Du, Kaichao Liang, and Yuxiang Xing.
\newblock Reduction of metal artefacts in ct with cycle-gan.
\newblock In {\em 2018 IEEE Nuclear Science Symposium and Medical Imaging Conference Proceedings (NSS/MIC)}, pages 1--3. IEEE, 2018.

\bibitem{yang2018unpaired}
Heran Yang, Jian Sun, Aaron Carass, Can Zhao, Junghoon Lee, Zongben Xu, and Jerry Prince.
\newblock Unpaired brain mr-to-ct synthesis using a structure-constrained cyclegan.
\newblock In {\em Deep Learning in Medical Image Analysis and Multimodal Learning for Clinical Decision Support: 4th International Workshop, DLMIA 2018, and 8th International Workshop, ML-CDS 2018, Held in Conjunction with MICCAI 2018, Granada, Spain, September 20, 2018, Proceedings 4}, pages 174--182. Springer, 2018.

\bibitem{liu2021ct}
Yanxia Liu, Anni Chen, Hongyu Shi, Sijuan Huang, Wanjia Zheng, Zhiqiang Liu, Qin Zhang, and Xin Yang.
\newblock Ct synthesis from mri using multi-cycle gan for head-and-neck radiation therapy.
\newblock {\em Computerized medical imaging and graphics}, 91:101953, 2021.

\bibitem{harms2019paired}
Joseph Harms, Yang Lei, Tonghe Wang, Rongxiao Zhang, Jun Zhou, Xiangyang Tang, Walter~J Curran, Tian Liu, and Xiaofeng Yang.
\newblock Paired cycle-gan-based image correction for quantitative cone-beam computed tomography.
\newblock {\em Medical physics}, 46(9):3998--4009, 2019.

\bibitem{karras2019style}
Tero Karras, Samuli Laine, and Timo Aila.
\newblock A style-based generator architecture for generative adversarial networks.
\newblock In {\em Proceedings of the IEEE/CVF conference on computer vision and pattern recognition}, pages 4401--4410, 2019.

\bibitem{fetty2020latent}
Lukas Fetty, Mikael Bylund, Peter Kuess, Gerd Heilemann, Tufve Nyholm, Dietmar Georg, and Tommy L{\"o}fstedt.
\newblock Latent space manipulation for high-resolution medical image synthesis via the stylegan.
\newblock {\em Zeitschrift f{\"u}r Medizinische Physik}, 30(4):305--314, 2020.

\bibitem{su2020pre}
Kang Su, Erning Zhou, Xiaoyu Sun, Che Wang, Dan Yu, and Xianlu Luo.
\newblock Pre-trained stylegan based data augmentation for small sample brain ct motion artifacts detection.
\newblock In {\em International Conference on Advanced Data Mining and Applications}, pages 339--346. Springer, 2020.

\bibitem{hong20213d}
Sungmin Hong, Razvan Marinescu, Adrian~V Dalca, Anna~K Bonkhoff, Martin Bretzner, Natalia~S Rost, and Polina Golland.
\newblock 3d-stylegan: A style-based generative adversarial network for generative modeling of three-dimensional medical images.
\newblock In {\em Deep Generative Models, and Data Augmentation, Labelling, and Imperfections: First Workshop, DGM4MICCAI 2021, and First Workshop, DALI 2021, Held in Conjunction with MICCAI 2021, Strasbourg, France, October 1, 2021, Proceedings 1}, pages 24--34. Springer, 2021.

\bibitem{mahapatra2019image}
Dwarikanath Mahapatra, Behzad Bozorgtabar, and Rahil Garnavi.
\newblock Image super-resolution using progressive generative adversarial networks for medical image analysis.
\newblock {\em Computerized Medical Imaging and Graphics}, 71:30--39, 2019.

\bibitem{upadhyay2021uncertainty}
Uddeshya Upadhyay, Yanbei Chen, Tobias Hepp, Sergios Gatidis, and Zeynep Akata.
\newblock Uncertainty-guided progressive gans for medical image translation.
\newblock In {\em Medical Image Computing and Computer Assisted Intervention--MICCAI 2021: 24th International Conference, Strasbourg, France, September 27--October 1, 2021, Proceedings, Part III 24}, pages 614--624. Springer, 2021.

\bibitem{armanious2020medgan}
Karim Armanious, Chenming Jiang, Marc Fischer, Thomas K{\"u}stner, Tobias Hepp, Konstantin Nikolaou, Sergios Gatidis, and Bin Yang.
\newblock Medgan: Medical image translation using gans.
\newblock {\em Computerized medical imaging and graphics}, 79:101684, 2020.

\bibitem{kingma2013auto}
Diederik~P Kingma and Max Welling.
\newblock Auto-encoding variational bayes.
\newblock {\em arXiv preprint arXiv:1312.6114}, 2013.

\bibitem{vahdat2020nvae}
Arash Vahdat and Jan Kautz.
\newblock Nvae: A deep hierarchical variational autoencoder.
\newblock {\em Advances in neural information processing systems}, 33:19667--19679, 2020.

\bibitem{hung2021hierarchical}
Alex Ling~Yu Hung, Zhiqing Sun, Wanwen Chen, and John Galeotti.
\newblock Hierarchical probabilistic ultrasound image inpainting via variational inference.
\newblock In {\em Deep Generative Models, and Data Augmentation, Labelling, and Imperfections: First Workshop, DGM4MICCAI 2021, and First Workshop, DALI 2021, Held in Conjunction with MICCAI 2021, Strasbourg, France, October 1, 2021, Proceedings 1}, pages 83--92. Springer, 2021.

\bibitem{cui2021pet}
Jianan Cui, Yutong Xie, Kuang Gong, Kyungsang Kim, Jaewon Yang, Peder Larson, Thomas Hope, Spencer Behr, Youngho Seo, Huafeng Liu, et~al.
\newblock Pet denoising and uncertainty estimation based on nvae model.
\newblock In {\em 2021 IEEE Nuclear Science Symposium and Medical Imaging Conference (NSS/MIC)}, pages 1--3. IEEE, 2021.

\bibitem{grover2020alignflow}
Aditya Grover, Christopher Chute, Rui Shu, Zhangjie Cao, and Stefano Ermon.
\newblock Alignflow: Cycle consistent learning from multiple domains via normalizing flows.
\newblock In {\em Proceedings of the AAAI Conference on Artificial Intelligence}, volume~34, pages 4028--4035, 2020.

\bibitem{bui2020flow}
Toan~Duc Bui, Manh Nguyen, Ngan Le, and Khoa Luu.
\newblock Flow-based deformation guidance for unpaired multi-contrast mri image-to-image translation.
\newblock In {\em Medical Image Computing and Computer Assisted Intervention--MICCAI 2020: 23rd International Conference, Lima, Peru, October 4--8, 2020, Proceedings, Part II 23}, pages 728--737. Springer, 2020.

\bibitem{wang2021harmonization}
Rongguang Wang, Pratik Chaudhari, and Christos Davatzikos.
\newblock Harmonization with flow-based causal inference.
\newblock In {\em Medical Image Computing and Computer Assisted Intervention--MICCAI 2021: 24th International Conference, Strasbourg, France, September 27--October 1, 2021, Proceedings, Part III 24}, pages 181--190. Springer, 2021.

\bibitem{hung2023med}
Alex Ling~Yu Hung, Kai Zhao, Haoxin Zheng, Ran Yan, Steven~S Raman, Demetri Terzopoulos, and Kyunghyun Sung.
\newblock Med-cdiff: Conditional medical image generation with diffusion models.
\newblock {\em Bioengineering}, 10(11):1258, 2023.

\bibitem{ho2020denoising}
Jonathan Ho, Ajay Jain, and Pieter Abbeel.
\newblock Denoising diffusion probabilistic models.
\newblock {\em Advances in neural information processing systems}, 33:6840--6851, 2020.

\bibitem{dhariwal2021diffusion}
Prafulla Dhariwal and Alexander Nichol.
\newblock Diffusion models beat gans on image synthesis.
\newblock {\em Advances in neural information processing systems}, 34:8780--8794, 2021.

\bibitem{saharia2022image}
Chitwan Saharia, Jonathan Ho, William Chan, Tim Salimans, David~J Fleet, and Mohammad Norouzi.
\newblock Image super-resolution via iterative refinement.
\newblock {\em IEEE transactions on pattern analysis and machine intelligence}, 45(4):4713--4726, 2022.

\bibitem{kadkhodaie2020solving}
Zahra Kadkhodaie and Eero~P Simoncelli.
\newblock Solving linear inverse problems using the prior implicit in a denoiser.
\newblock {\em arXiv preprint arXiv:2007.13640}, 2020.

\bibitem{meng2021sdedit}
Chenlin Meng, Yutong He, Yang Song, Jiaming Song, Jiajun Wu, Jun-Yan Zhu, and Stefano Ermon.
\newblock Sdedit: Guided image synthesis and editing with stochastic differential equations.
\newblock {\em arXiv preprint arXiv:2108.01073}, 2021.

\bibitem{sinha2021d2c}
Abhishek Sinha, Jiaming Song, Chenlin Meng, and Stefano Ermon.
\newblock D2c: Diffusion-decoding models for few-shot conditional generation.
\newblock {\em Advances in Neural Information Processing Systems}, 34:12533--12548, 2021.

\bibitem{sasaki2021unit}
Hiroshi Sasaki, Chris~G Willcocks, and Toby~P Breckon.
\newblock Unit-ddpm: Unpaired image translation with denoising diffusion probabilistic models.
\newblock {\em arXiv preprint arXiv:2104.05358}, 2021.

\bibitem{saharia2022palette}
Chitwan Saharia, William Chan, Huiwen Chang, Chris Lee, Jonathan Ho, Tim Salimans, David Fleet, and Mohammad Norouzi.
\newblock Palette: Image-to-image diffusion models.
\newblock In {\em ACM SIGGRAPH 2022 conference proceedings}, pages 1--10, 2022.

\bibitem{behrendt2024patched}
Finn Behrendt, Debayan Bhattacharya, Julia Kr{\"u}ger, Roland Opfer, and Alexander Schlaefer.
\newblock Patched diffusion models for unsupervised anomaly detection in brain mri.
\newblock In {\em Medical Imaging with Deep Learning}, pages 1019--1032. PMLR, 2024.

\bibitem{rahman2023ambiguous}
Aimon Rahman, Jeya Maria~Jose Valanarasu, Ilker Hacihaliloglu, and Vishal~M Patel.
\newblock Ambiguous medical image segmentation using diffusion models.
\newblock In {\em Proceedings of the IEEE/CVF Conference on Computer Vision and Pattern Recognition}, pages 11536--11546, 2023.

\bibitem{wu2024medsegdiff}
Junde Wu, Rao Fu, Huihui Fang, Yu~Zhang, Yehui Yang, Haoyi Xiong, Huiying Liu, and Yanwu Xu.
\newblock Medsegdiff: Medical image segmentation with diffusion probabilistic model.
\newblock In {\em Medical Imaging with Deep Learning}, pages 1623--1639. PMLR, 2024.

\bibitem{zbinden2023stochastic}
Lukas Zbinden, Lars Doorenbos, Theodoros Pissas, Adrian~Thomas Huber, Raphael Sznitman, and Pablo M{\'a}rquez-Neila.
\newblock Stochastic segmentation with conditional categorical diffusion models.
\newblock In {\em Proceedings of the IEEE/CVF International Conference on Computer Vision}, pages 1119--1129, 2023.

\bibitem{chen2023berdiff}
Tao Chen, Chenhui Wang, and Hongming Shan.
\newblock Berdiff: Conditional bernoulli diffusion model for medical image segmentation.
\newblock In {\em International Conference on Medical Image Computing and Computer-Assisted Intervention}, pages 491--501. Springer, 2023.

\bibitem{cao2024high}
Chentao Cao, Zhuo-Xu Cui, Yue Wang, Shaonan Liu, Taijin Chen, Hairong Zheng, Dong Liang, and Yanjie Zhu.
\newblock High-frequency space diffusion model for accelerated mri.
\newblock {\em IEEE Transactions on Medical Imaging}, 2024.

\bibitem{luo2023bayesian}
Guanxiong Luo, Moritz Blumenthal, Martin Heide, and Martin Uecker.
\newblock Bayesian mri reconstruction with joint uncertainty estimation using diffusion models.
\newblock {\em Magnetic Resonance in Medicine}, 90(1):295--311, 2023.

\bibitem{xie2022measurement}
Yutong Xie and Quanzheng Li.
\newblock Measurement-conditioned denoising diffusion probabilistic model for under-sampled medical image reconstruction.
\newblock In {\em International Conference on Medical Image Computing and Computer-Assisted Intervention}, pages 655--664. Springer, 2022.

\bibitem{dinh2016density}
Laurent Dinh, Jascha Sohl-Dickstein, and Samy Bengio.
\newblock Density estimation using real nvp.
\newblock {\em arXiv preprint arXiv:1605.08803}, 2016.

\bibitem{xiao2021tackling}
Zhisheng Xiao, Karsten Kreis, and Arash Vahdat.
\newblock Tackling the generative learning trilemma with denoising diffusion gans.
\newblock {\em arXiv preprint arXiv:2112.07804}, 2021.

\bibitem{wiatrak2019stabilizing}
Maciej Wiatrak, Stefano~V Albrecht, and Andrew Nystrom.
\newblock Stabilizing generative adversarial networks: A survey.
\newblock {\em arXiv preprint arXiv:1910.00927}, 2019.

\bibitem{miyato2018spectral}
Takeru Miyato, Toshiki Kataoka, Masanori Koyama, and Yuichi Yoshida.
\newblock Spectral normalization for generative adversarial networks.
\newblock {\em arXiv preprint arXiv:1802.05957}, 2018.

\bibitem{motwani2020novel}
Tanya Motwani and Manojkumar Parmar.
\newblock A novel framework for selection of gans for an application.
\newblock {\em arXiv preprint arXiv:2002.08641}, 2020.

\bibitem{davidson2018hyperspherical}
Tim~R Davidson, Luca Falorsi, Nicola De~Cao, Thomas Kipf, and Jakub~M Tomczak.
\newblock Hyperspherical variational auto-encoders.
\newblock {\em arXiv preprint arXiv:1804.00891}, 2018.

\bibitem{asperti2019variational}
Andrea Asperti et~al.
\newblock Variational autoencoders and the variable collapse phenomenon.
\newblock {\em Sensors \& Transducers}, 234(6):1--8, 2019.

\bibitem{montecchiani2022survey}
Christian Montecchiani.
\newblock Survey of modern generative modelling.
\newblock {\em Science (CS-E4000), Fall 2022}, 2022.

\bibitem{kazerouni2023diffusion}
Amirhossein Kazerouni, Ehsan~Khodapanah Aghdam, Moein Heidari, Reza Azad, Mohsen Fayyaz, Ilker Hacihaliloglu, and Dorit Merhof.
\newblock Diffusion models in medical imaging: A comprehensive survey.
\newblock {\em Medical Image Analysis}, page 102846, 2023.

\bibitem{sohl2015deep}
Jascha Sohl-Dickstein, Eric Weiss, Niru Maheswaranathan, and Surya Ganguli.
\newblock Deep unsupervised learning using nonequilibrium thermodynamics.
\newblock In {\em International conference on machine learning}, pages 2256--2265. PMLR, 2015.

\bibitem{rink2021memory}
Norman~A Rink, Adam Paszke, Dimitrios Vytiniotis, and Georg~Stefan Schmid.
\newblock Memory-efficient array redistribution through portable collective communication.
\newblock {\em arXiv preprint arXiv:2112.01075}, 2021.

\bibitem{song2019generative}
Yang Song and Stefano Ermon.
\newblock Generative modeling by estimating gradients of the data distribution.
\newblock {\em Advances in neural information processing systems}, 32, 2019.

\end{thebibliography}

\end{document}